\documentclass[aps,pre,twocolumn,showpacs,showkeys,floatfix]{revtex4}

\usepackage[dvips]{graphicx}
\usepackage{amsmath}

\newcommand{\xm}{\langle x \rangle}
\newcommand{\ym}{\langle y \rangle}
\newcommand{\mean}[1]{\langle #1 \rangle}

\begin{document}
\title{Macroscopic limit cycle via pure noise-induced phase transition}
\author{R. Kawai}
\email{kawai@uab.edu}
\homepage{http://kawai.phy.uab.edu/}
\affiliation{Department of Physics, University of Alabama at Birmingham,
1300 University Blvd., Birmingham, AL 35294, USA} 
\author{X. Sailer}
\author{L. \surname{Schimansky-Geier}}
\affiliation{Institute of Physics, Humboldt University at Berlin,
Newtonstra{\ss}e 15, 12489l Berlin, Germany}
\author{C. \surname{Van den Broeck}}
\affiliation{Department WNI, Limburgs Universitaire Centrum, 
Universitaire Campus B-3590 Diepenbeek, Belgium}

\begin{abstract}
Bistability generated via a pure noise-induced phase transition is
reexamined from the view of bifurcations in macroscopic cumulant
dynamics. It allows an analytical study of the phase diagram in 
more general cases than previous methods. In
addition using this approach we investigate 
spatially-extended systems with two degrees of freedom per site. For
this system, the analytic solution of the stationary Fokker-Planck
equation is not available and a standard mean field approach cannot be
used to find noise induced phase
transitions. A new approach based on cumulant dynamics predicts
a noise-induced phase transition 
through a Hopf bifurcation leading to a macroscopic limit cycle
motion, which is confirmed by numerical simulation.
\end{abstract}

\date{\today}


\maketitle 

\section{INTRODUCTION}
\label{sec:intro}
The interplay between nonlinear dynamics and noise often generates
interesting and counterintuitive phenomena. Popular examples are
stochastic resonance \cite{gammaitoni98}, coherence resonance 
\cite{anishchenko02}, and noise induced phase transitions
\cite{vandenbroeck94,vandenbroeck97,jordi99}. In the latter 
example noise creates an ordered phase which does not exist in the
absence of noise. Unlike noise induced transitions in systems with few
degrees of freedom, the noise induced phase transition breaks
ergodicity and has the characteristics of a genuine phase transition.
In previous studies, many variations of pure noise-induced phase
transitions were introduced \cite{jordi99}. Spatial patterns can be
induced via the pure noise-induced phase transition
\cite{jordi93,parrondo96,zaikin98,buceta03}. The noise-induced first order phase
transition was also shown to be possible
\cite{muller97,zaikin99,kim98}. The systems with colored noise  
were investigated by various groups \cite{jordi92,kim98,mangioni97,mangioni00}.
Furthermore, the bistability created by the noise-induced phase
transition exhibits stochastic resonance when a time-periodic
perturbation is added \cite{zaikin00} and it can lead to propagation
of harmonic signals \cite{zaikin02}.  The idea of noise-induced phase
transition was also used in coupled Brownain motors. \cite{reimann99a,reimann99b}

Most of previous investigations take a mean field approach and use
the self-consistent condition \cite{jordi99,shiino87}
\begin{equation}
\label{eq:SelfCon}
\xm = \int_{-\infty}^{\infty} x P^{st}\left(x;\xm \right) dx.
\end{equation}
to determine the mean $\xm$ and also bifurcation points.
This method yields an exact solution for the phase boundaries.
However, to solve Eq.~(\ref{eq:SelfCon}) one needs to know the analytic
solution of the stationary Fokker-Planck equation
$P^{st}\left(x;\xm\right)$, which is in general not available. 
Furthermore, if the system does not have a stationary state (i.~e.,
$\xm$ is time-dependent), we must use time-dependent self-consistent condition
which is prohibitively more difficult.
For certain types of spatially extended stochastic problems,
there is an approximate method which replaces stochastic
dynamics with effective deterministic dynamics \cite{santos01}.  However, 
the extent of applicability to other models is not known.
A general and systematic method is highly desired.

In this paper we present a systematic
method to investigate noise-induced phase transitions.  While it does
not provide exact solutions, the method does 
not require an analytical expression of the stationary state probability
distribution and can be applied to general cases including
time-dependent problems.  In the following section, we investigate a
model system with a single variable introduced by Van den Broeck {\it
et~al.} \cite{vandenbroeck94}  
for which an exact solution is known.  The present method predicts a
pitchfork bifurcation to an ordered state as the noise intensity
increases and also the reentrant transition to a disordered phase at
a higher noise intensity.  This behavior is qualitatively in a good
agreement with the exact results. 

Then, we apply the same method to a
model with two variables.  The model is expected to undergo a
noise-induced phase transition to a time-dependent ordered phase for
which the time-dependent self-consistent approach is not practical.
The present method
predicts a Hopf bifurcation to a macroscopic limit cycle phase from a
disordered state as the noise intensity increases and shows also a
reentrant 
transition.  We also demonstrate that the present method can provide
other information such as a period and amplitude of the oscillation.

\section{NOISE-INDUCED PITCHFORK BIFURCATION}
\label{sec:pitchfork}

In this section, we consider the following stochastic system of
$N$ globally coupled microscopic variables $\{x_i\}$:
\begin{equation}
\label{eq:Langevin1}
\dot{x_i} = f(x_i) - \frac{D}{N} \sum_{j=1}^N  (x_i-x_j) + g(x_i)
\xi_i(t) 
\end{equation}
where $D$ is a coupling strength and 
$\xi_i(t)$ is a Gaussian white noise,
defined by
\begin{equation}
\label{eq:noise}
\mean{\xi_i(t)} = 0, \qquad
\mean{\xi_i(t)\xi_j(t')}=\sigma^2\delta_{ij}\delta(t-t').
\end{equation}
Equation (\ref{eq:Langevin1}) is interpreted in the Stratonovich sense.

With certain nonlinear functions $f(x)$ and $g(x)$,
Eq.~(\ref{eq:Langevin1}) exhibits a 
phase transition from a disordered phase ($\xm = 0$) to
an ordered phase ($\xm \ne 0$) as the noise
intensity $\sigma^2$ increases. At larger noise intensities the
system undergoes a reentrant transition to another disordered phase
($\xm = 0$) \cite{vandenbroeck94,vandenbroeck97}.
Since the ordered phase does not exist in the absence of noise, the
phenomena is called {\it pure} noise-induced phase transition.
Van~den~Broeck {\it et~al.} \cite{vandenbroeck94} originally
investigated it using the following nonlinear functions:
\begin{equation}
f(x) = -x (1+x^2)^2, \qquad 
g(x) = 1+x^2
\label{eq:FandG}
\end{equation}
because a stationary state solution to the corresponding Fokker-Planck
equation can be obtained analytically, which allowed them to
find the exact phase boundary with the self-consistent equation
(\ref{eq:SelfCon}).  

In the following, we present a method which allows us to investigate
more general cases approximately but without an analytical probability
distribution.  We apply the method to this model
(\ref{eq:FandG}) and compare the results with the exact solution and
also with numerical simulations.

\subsection{Moment Dynamics}
\label{subsec:MomentDyn1}

Assuming $N \rightarrow \infty$, we write Eq.~(\ref{eq:Langevin1}) in a
mean field form:
\begin{equation}
\label{eq:LangevinMF}
\dot{x} = f(x) - D (x-\xm) + g(x)\xi(t) .
\end{equation}
Taking the mean of Eq.~(\ref{eq:LangevinMF}) under the Stratonovich
interpretation, the dynamics of $\xm$ is given by
\begin{equation}
\label{eq:MfDyn1}
\dot{\mean{x}} = \mean{f(x)} + \frac{\sigma^2}{2} \mean{g'(x)g(x)} .
\end{equation}
Expanding $f(x)$ and $g(x)$ in Taylor's series around $\xm$,  
Eq.~(\ref{eq:MfDyn1}) forms an infinite set of simultaneous
ordinary differential equations:
\begin{equation}
  \dot{\xm} = \sum_{n=0} \frac{\mu_n}{n!} \left \{f^{(n)}(\xm) 
        + \frac{\sigma^2}{2}[g'(\xm)g(\xm)]^{(n)}\right\}
        \label{eq:xmDot1}
\end{equation}
\begin{eqnarray}
\dot{\mu}_n &=& -D n \mu_n \nonumber \\
        &+& \sum_{m=0} \frac{ n \mu_{n+m-1}}{m!}
        \left \{
        f^{(m)}(\xm) + \frac{\sigma^2}{2}[g'(\xm)g(\xm)]^{(m)}
        \right \} \nonumber \\
        &+& \sum_{m=0} \frac{n(n-1) \mu_{n+m-2}}{m!} \frac{\sigma^2}{2}
        [g^2(\xm)]^{(m)} .
        \label{eq:muDot1}
\end{eqnarray}
Here, $f^{(n)}$ is the $n$-th order derivative and $\mu_n =
\mean{(x-\xm)^n}$ the $n$-th central moment with, by definition,
$\mu_0=1$ and $\mu_1=0$. 

Some previous studies \cite{vandenbroeck97,GarSag02} have
considered only the 0-th order term, 
thereby
neglecting the fluctuations $x-\xm$, and worked with the equation
\begin{equation}
\label{eq:ZerothOrder}
\mean{\dot{x}}=f(\xm)+\frac{\sigma^2}{2}g'(\xm)g(\xm) .
\end{equation}
Since this equation does not depend on the coupling constant $D$,
it cannot explain the reentrant transition.  In fact,
Eq.~(\ref{eq:ZerothOrder}) is 
exact when $D \rightarrow \infty$ under which condition no reentrant
transition 
takes place. \cite{vandenbroeck94,vandenbroeck97} 
For a finite $D$, one must consider higher order terms.

In order to see how the higher order terms create the reentrant
transition, we investigate the previous model Eq. (\ref{eq:FandG}).
Here we show the equations of motion only for $\xm$ and $\mu_2$:
\begin{eqnarray}
  \mean{\dot{x}}
        &=& (\sigma^2 -2 ) \mu_3 - \mu_5 \nonumber \\
        &+& \left [ \sigma^2 -1 + 3(\sigma^2- 2)\mu_2 
        - 5\mu_4 \right] \xm 
         \nonumber \\
        &-& 10\mu_3 \xm^2 + (\sigma^2-2-10\mu_2) \xm^3 - \xm^5 
        \label{eq:xmDot2}
\end{eqnarray}
\begin{eqnarray}
  \dot{\mu_2} 
        &=& \sigma^2 + 2[2 \sigma^2 - (1+D)]\mu_2 
        + \left ( 3\sigma^2 - 4\right ) \mu_4 - 2\mu_6
        \nonumber \\
        &+& \left [ 2 (5\sigma^2 -6)\mu_3 - 10 \mu_5 \right ] \xm
        \nonumber \\
        &+& \left [2\sigma^2+12(\sigma^2-1)\mu_2-20\mu_4 \right ]
\xm^2 
        \nonumber \\
        &-& 20\mu_3 \xm^3 + (\sigma^2-10\mu_2)\xm^4 .
        \label{eq:mu2Dot2}
\end{eqnarray}
Even with higher order terms,  Eq.~(\ref{eq:xmDot2}) still does not
explicitly 
depend on the coupling constant.  The effect of coupling arises from
the dynamics of the second and higher order moments as
Eq.~(\ref{eq:mu2Dot2}) indicates.

Noting $f(-x)=-f(x)$ and $g(-x)=g(x)$, the system
(\ref{eq:LangevinMF})
is invariant 
under the variable transformation $x \rightarrow -x$.  Due to this
symmetry the odd 
moments 
must be zero when $\xm=0$.  Then, we find a fixed point at
$\xm^*=0$ and $\mu^*_{2n+1}=0$.  The even moments at this
fixed point are not
zero and must be
determined by setting the r.h.s. of Eq.~(\ref{eq:muDot1}) to zero. For example,
Eq.~(\ref{eq:mu2Dot2}) provides the following equation:
\begin{equation}
\sigma^2 + 2[2 \sigma^2 - (1+D)]\mu^*_2 
+ \left ( 3\sigma^2 - 4 \right ) \mu^*_4 - 2\mu^*_6 .
= 0
\label{eq:EvenMoments}
\end{equation}
%
\begin{figure}[t]
\begin{center}
\includegraphics[width=3.3in]{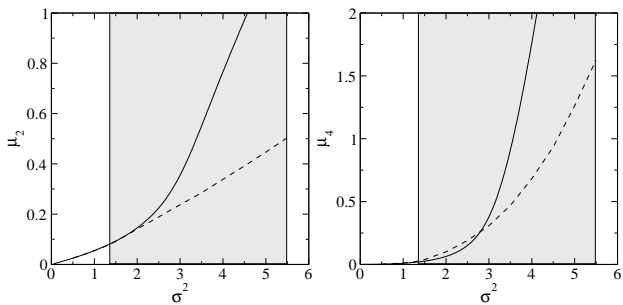} 
\caption{Stationary moments $\mu^*_{2}$ (left panel) and $\mu^*_{4}$
(right panel) obtained by Gaussian
approximation (solid lines) and numerical
simulation (dashed lines).  The coupling strength $D=10$ is used.
The Gaussian approximation appears in a good agreement with the
simulation below $\sigma^2=2$.  However, it overestimates both 2nd and
4th order moments above $\sigma^2=2$. This rapid growth causes
reentrant transition at a smaller noise intensity than the exact
solution. The grey region shows the location of the ordered phase.
}
\label{fig:Moments}
\end{center}
\end{figure}

A linear stability analysis of Eq.~(\ref{eq:xmDot2}) yields the
bifurcation condition for a pitchfork bifurcation:
\begin{equation}
\label{eq:ExactBifurcation1}
\sigma_c^2 -1 + 3(\sigma_c^2- 2)\mu^*_2 - 5\mu^*_4  = 0
\end{equation}
where $\sigma_c$ is a critical noise intensity.
Even without the exact knowledge of the higher moments
useful information can be derived from
Eq.~(\ref{eq:ExactBifurcation1}). Since 
both $\mu_2$ and $\mu_4$ are non-negative, l.h.s. is alway negative
for $\sigma^2<1$. In this regime the fixed point $\xm^*=0$ is
stable regardless of the magnitude of $D$. 
For $1<\sigma^2<2$, only the 0-th moment term is positive but
the 
other terms are negative and support stability.
The bifurcation is possible  in this range of noise intensity 
only when the moments are sufficiently
small.  Since increasing the coupling strength
reduces fluctuation, the bifurcation takes place above a certain
magnitude of $D$.

Interestingly, the role of the second moment term changes at
$\sigma^2=2$. At higher values it supports instability of the fixed
point.  The 4-th moment term
is always negative and supports stability. When $\mu_4$ grows faster
than $\mu_2$ with 
increasing $\sigma^2$, it eventually dominates and l.h.s. of
Eq.~(\ref{eq:ExactBifurcation1}) becomes negative again.  
Then, the system reenters the  disordered phase. 
In order to determine the critical noise intensity $\sigma_c$ from
Eq.~(\ref{eq:ExactBifurcation1}), one needs to know stationary
even moments $\mu^*_2$ and $\mu^*_4$, which are to be determined by
Eq.~(\ref{eq:EvenMoments}).  In turn, it requires $\mu^*_6$.
At the end, all even moments must be
simultaneously solved, which is practically impossible for general
cases. An approximation is necessary.

\subsection{Gaussian Approximation}
\label{gauss}

\begin{figure}
\begin{center}
\rotatebox{270}{
\includegraphics[width=2.2in]{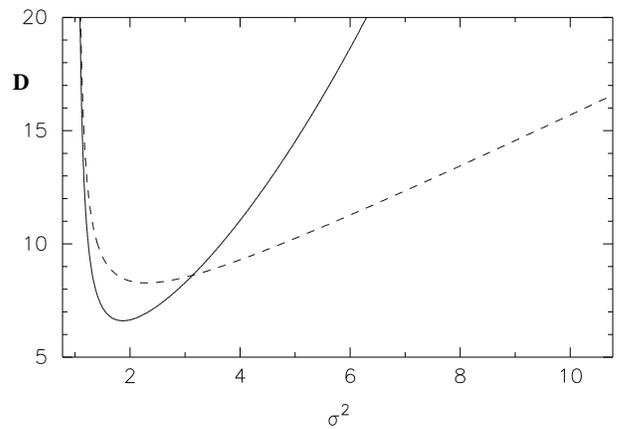}}
\caption{The phase diagram. The ordered phase ($\xm \ne 0$) is above
the lines 
and the disordered phase ($\xm=0$) below them. The dashed line is
the exact solution obtained by Eq. \ref{eq:SelfCon}, the solid line
is   the Gaussian approximation.} 
\label{fig:PitchforkPhase}
\end{center}
\end{figure}

When $f(x)$ or $g(x)$ is nonlinear, the stochastic dynamics
(\ref{eq:Langevin1}) is not a Gaussian process and in general
we cannot solve the system of equations (\ref{eq:xmDot1}) and
(\ref{eq:muDot1}) exactly. For
an approximate solution it is convenient to assume a probability
distribution $P(x;\xm)$ for which the cumulants above a certain
order are negligible. We chose here the simplest example where the
cumulants above the second order are set to zero (Gaussian
approximation). In general, this approximation is quantitatively
not justified
but reproduces the main features of the noise-induced phase transition,
especially the reentrant transition into the disordered phase.

In the Gaussian approximation all odd moments vanish. The even moments
can be expressed by the 2nd moment $\mu_2$. 
For example,
\begin{equation}
\mu_4=3 \mu_2^2 \quad \text{and} \quad \mu_6=15 \mu_2^3 .
\label{eq:GaussianMoments}
\end{equation}
Applying these relations, Eq.~(\ref{eq:EvenMoments}) becomes
\begin{equation}
\label{eq:GaussianMu2}
\sigma^2 + 2[2 \sigma^2 - (1+D)]\mu_2 
+ 3 \left ( 3 \sigma^2 - 4 \right ) (\mu^*_2)^2 - 30(\mu^*_2)^3 = 0
\end{equation}
which determines $\mu^*_2$ as a function of $\sigma^2$.  In turn, we can
determine higher order even moments via the Gaussian approximation. 
Figure \ref{fig:Moments} plots $\mu^*_2$ and $\mu^*_4$ obtained by 
the Gaussian approximation and also the results of numerical
simulation 
for comparison.  Up to $\sigma^2=2$, the Gaussian
approximation is in a good agreement with the simulation.  At higher
noise intensities, non-Gaussian behavior becomes large and the Gaussian
approximation significantly overestimates the fluctuation.

Once we find all stationary moments, we can quantitatively evaluate
the stability condition (\ref{eq:ExactBifurcation1}) for $\xm$. With
the Gaussian approximation 
(\ref{eq:GaussianMoments}), the bifurcation condition
(\ref{eq:ExactBifurcation1}) 
becomes: 
\begin{equation}
\label{eq:GaussianBifurcation1}
\sigma_c^2 -1 + 3(\sigma_c^2- 2)\mu^*_2 - 15(\mu^*_2)^2 = 0 \quad
\end{equation}
Here, $\mu^*_2$ must also satisfy Eq. (\ref{eq:GaussianMu2}) with
noise intensity $\sigma_c$. In other words,
Eqs.~(\ref{eq:GaussianMu2}) and (\ref{eq:GaussianBifurcation1}) must be 
solved simultaneously for $\sigma_c$ and $\mu_2^*$.  

Figure~\ref{fig:PitchforkPhase} shows the results and compares them with
the exact solution obtained from 
Eq.~(\ref{eq:SelfCon}). Although there is a clear quantitative
discrepancy 
between the approximation and the exact solution the main features,
namely the entrant transition into the ordered phase at medium noise
intensities 
and the reentrant into the disordered phase at high intensities is
reproduced. As in the exact solution a certain minimum coupling strength is
required  for the transition to take place. 

\begin{figure}[t]
\begin{center}
\rotatebox{270}{
\includegraphics[width=2.2in,angle=90]{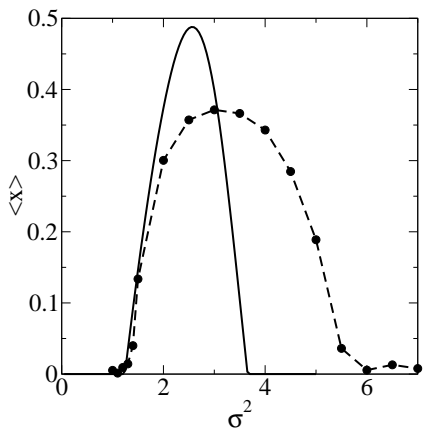}}
\caption{The mean $\xm$ by the Gaussian approximation (solid line) and
by numerical simulation (dashed line).  The coupling strength
$D$=10 is used.} 
\label{fig:xMean}
\end{center}
\end{figure}

We can also evaluate the mean $\xm$ as a function of $\sigma^2$ from
Eqs.~(\ref{eq:xmDot2}), (\ref{eq:GaussianMoments}), and
(\ref{eq:GaussianMu2}). 
The result is plotted in Fig. \ref{fig:xMean} along with the results of
simulation.  The agreement near the first transition point is quite
good. 
However, the Gaussian approximation predicts the reentrant
transition much earlier than the simulation as mentioned earlier.
%

We noticed that the present results resemble the
phase boundary for the 2-dimensional system with a local coupling
examined in Ref. \cite{vandenbroeck97} surprisingly well. This
coincidence is partly 
due to the fact that the locally coupled
system has much larger fluctuation than the globally
coupled system, which induces the reentrant transition at a smaller
noise intensity. Since the Gaussian approximation overestimates the
fluctuation, it shares some similarity with locally coupled systems.

\section{Noise-Induced Limit Cycle}
\label{limitCycle}

\begin{figure}[t]
\begin{center}
\includegraphics[width=2.5in]{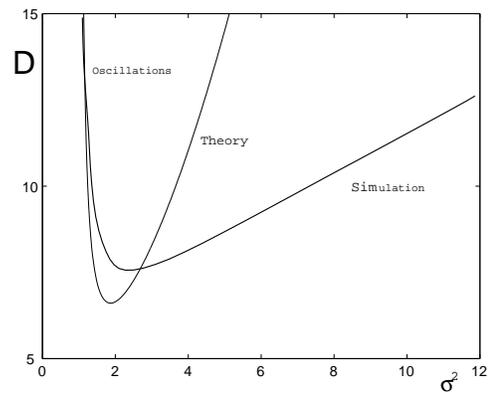}
\caption{Phase diagram for the two variable case. In the region 
above the lines we
observe oscillations of the mean, below it we do not. The line labeled
theory is the analytic solution in the Gaussian approximation, the
line labeled simulation is the numerically obtained result for
the Langevin dynamics of Eq.~\ref{eq:Langevin2} and \ref{eq:sdex} for
$N=625$ systems.} 
\label{fig:HopfPhase}
\end{center}
\end{figure}

In this section we investigate a model with two variables at each
site.  
Multivariate stochastic systems are mathematically quite difficult.
Even if the system has a stationary state, it is usually hard to
find an analytical expression of its probability distribution. It
will be even more difficult if the system does not have a stationary
state and the probability distribution is explicitly time-dependent.
The lack of an analytical expression of the probability distribution
makes the standard method based on the self-consistent
equation (\ref{eq:SelfCon}) 
futile. Direct numerical simulation of multivariate Langevin
equations demands more computational power than the single variable
cases.  However, we expect that the Gaussian approximation provides a
degree of accuracy similar to that in the single variable case without
increasing mathematical difficulty.

We use the following simple model that keeps a close connection to the
previous 
model:
\begin{eqnarray}
\label{eq:Langevin2}
\dot{x_i} &=& f(x_i) - \frac{D}{N}\sum_j (x_i-x_j) + g(x_i)\xi_i(t) -
y_i\\
\dot{y_i} &=& k x_i  \label{eq:sdex} 
\end{eqnarray}
with $f(x)$ and $g(x)$ defined by Eq.~(\ref{eq:FandG}) as before. 
Again, we expand the dynamical equations in terms of the central
moments:
\begin{eqnarray}
  \mean{\dot{x}} &=& \sum_{n=0} \frac{\mu_{n,0}}{n!} \left \{ 
        f^{(n)}(\xm) + \frac{\sigma^2}{2}[g'(\xm)g(\xm)]^{(n)}\right
        \}\nonumber \\
        &-& \ym
        \label{eq:xmDot3}
\end{eqnarray}
\begin{equation}
  \mean{\dot{y}} = k \xm 
        \label{eq:ymDot3}
\end{equation}
\begin{eqnarray}
  \dot{\mu}_{n,m} &=& -nD\mu_{n,m}-n\mu_{n-1,m+1} \nonumber \\
        &+& mk\mu_{n+1,m-1} + mk\mu_{n,m-1}\mean{x}
            -n\mu_{n-1,m}\mean{y} \nonumber \\
        &+& \sum_{\ell=0} \frac{ n \mu_{n+\ell-1,m}}{\ell!}
        \left \{
        f^{(\ell)}(\xm) + \frac{\sigma^2}{2}[g'(\xm)g(\xm)]^{(\ell)}
        \right \}\nonumber \\
        &+&\sum_{\ell=0} \frac{n(n-1) 
        \mu_{n+\ell-2,m}}{\ell!} \frac{\sigma^2}{2}
        [g^2(\xm)]^{(\ell)}
        \label{eq:muDot3}
\end{eqnarray}
where $n+m \ge 2$ and 
$\mu_{n,m}=\left\langle (x-<x>)^n (y-<y>)^m \right \rangle$ with 
$\mu_{0,0}=1$ and $\mu_{1,0}=\mu_{0,1}=0$.
%
\begin{figure}[tb]
\begin{center}
\includegraphics[width=3.in]{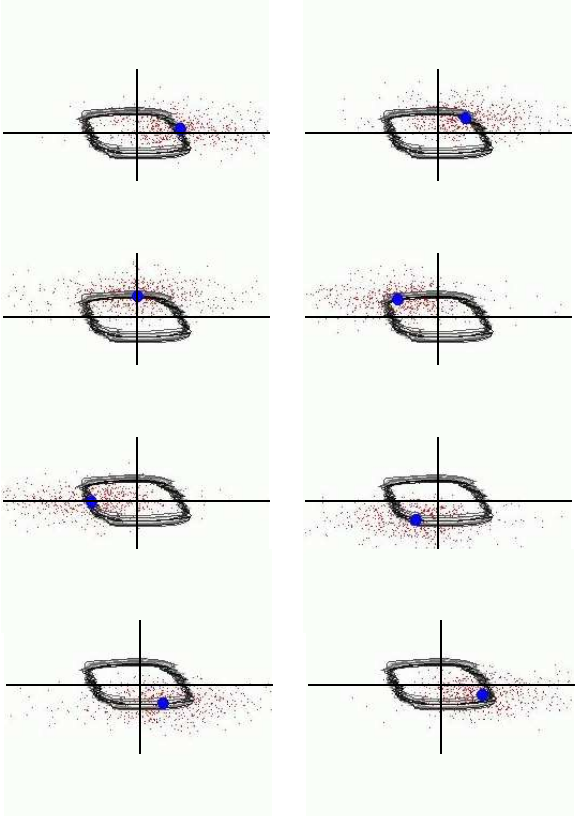}
\caption{Snapshots of an ensemble of 625 elements in phase-space. The
  single dots denote the individual elements. The solid circle shows the
  mean and the line depicts the trajectory of the mean. The time series
  goes from left to right and from the top to the bottom. Parameter
values $D=10$ and $k=0.1$ are used.} 
\label{fig:Simulation}
\end{center}
\end{figure}

For the present model (\ref{eq:FandG}), 
Eqs.~(\ref{eq:xmDot3}) and (\ref{eq:muDot3}) are explicitly written as
\begin{eqnarray}
\mean{\dot{x}} &=& (\sigma^2-2)\mu_{3,0}-\mu_{5,0} \nonumber \\*
        &+&\left [ \sigma^2-1+3(\sigma^2-2)\mu_{2,0}-5\mu_{4,0} \right
] \xm \nonumber \\*
        &-&10\mu_{3,0}\xm^2 +(\sigma^2-2-10\mu_{2,0})\xm^3 \nonumber \\*
        &-&\xm^5 -\ym \label{eq:xmDot4}
\end{eqnarray}
\begin{eqnarray}
\dot{\mu}_{2,0} 
        &=&\sigma^2 + 2\left[2 \sigma^2 - (1+D)\right]\mu_{2,0} \nonumber \\*
        &-&2\mu_{1,1} 
           +( 3\sigma^2-4 )\mu_{4,0} \nonumber \\*
        &-& 2\mu_{6,0} 
            + 2\left[(5\sigma^2-6)\mu_{3,0}-5\mu_{5,0}\right] \xm
            \nonumber \\*
        &+&2\left[\sigma^2+6(\sigma^2-1)\mu_{2,0}-10\mu_{4,0}\right]\xm^2
        \nonumber \\*
        &-&20\mu_{3,0}\xm^3 +(\sigma^2-10\mu_2)\xm^4
        \label{eq:mu20Dot4}
\\
\nonumber\\
\dot{\mu}_{0,2}
        &=&2k\mu_{1,1}  \label{eq:mu02Dot4}
\end{eqnarray}
\begin{eqnarray}\dot{\mu}_{1,1}
        &=&(\sigma^2-D-1)\mu_{1,1}+k\mu_{2,0}-\mu_{0,2}+(\sigma^2-2)\mu_{3,1}
        \nonumber \\*
        &-&\mu_{5,1}+\left[3(\sigma^2-2)\mu_{2,1}-5\mu_{4,1}\right]\xm
        \nonumber \\*
        &+&\left[3(\sigma^2-2)\mu_{1,1}-10\mu_{3,1}\right]\xm^2
        \nonumber \\*
        &-&10\mu_{2,1}\xm^3-5\mu_{1,1}\xm^4  .
        \label{eq:mu11Dot4}
\end{eqnarray}
Here only the lowest order moments are shown.

Taking into account the symmetry of the model,
there is at least one fixed point at $\xm^*=\ym^*=0$ and
$\mu^*_{n,2m+1}=\mu^*_{2m+1,n}=0$.  Stationary even moments are
determined by an infinite set of simultaneous equation.  Here we show
three conditions derived from
Eqs. (\ref{eq:mu20Dot4})- (\ref{eq:mu11Dot4}):
\begin{equation}
\sigma^2 + 2\left[2 \sigma^2 - (1+D)\right]\mu^*_{2,0} 
 +( 3\sigma^2-4 )\mu^*_{4,0} -2\mu^*_{6,0} = 0, 
\label{eq:EquilibriumMu20}
\end{equation}
\begin{equation}
k\mu^*_{2,0}-\mu^*_{0,2} + \left ( \sigma^2-2 \right)\mu^*_{3,1}
- \mu^*_{5,1}=0 
\label{eq:EquilibriumMu02}
\end{equation}
and $\mu^*_{1,1}=0$.

Now, we apply the Gaussian approximation to this system.  For any two
variable Gaussian system, all odd moments are zero ($\mu_{m,n}=0$
for $m+n=\text{odd integer}$) and any even oder moment can be expressed
as a product of the 2nd order moments, $\mu_{2,0}$, $\mu_{0,2}$, and
$\mu_{1,1}$. 
For the present model we need only the following relations:
\begin{eqnarray}
&&\mu_{3,1}=3\mu_{2,0}\mu_{1,1},\quad \mu_{4,0}=3\mu^2_{2,0} \nonumber\\*
&&\mu_{5,1}=15\mu^2_{2,0}\mu_{1,1},\quad \mu_{6,0}=15\mu^3_{2,0}  .
\label{eq:GaussianApprox2}
\end{eqnarray}
Under the Gaussian approximation, the stationary even moments
$\mu^*_{2,0}$ and $\mu^*_{0,2}$ can be determined by 
Eqs.~(\ref{eq:EquilibriumMu20}) and
(\ref{eq:EquilibriumMu02}).
%
\begin{figure}[t]
\begin{center}
\includegraphics[width=2.5in]{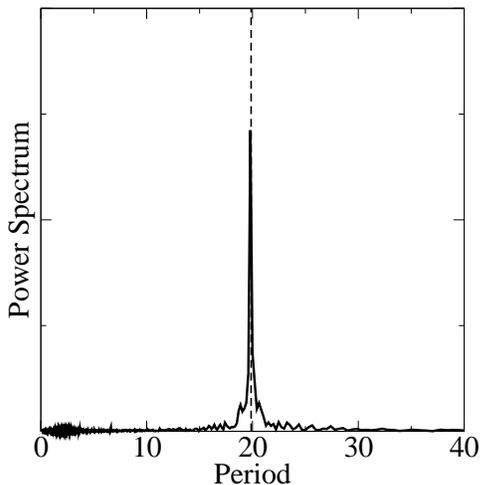}
\caption{Power spectrum of a limit cycle motion obtained by numerical
simulation (solid line) and the period obtained from the linear
stability analysis (dashed 
line).  Parameter values are $D$=10, $\sigma^2$=2, and $k$=0.1. The
agreement is perfect.} 
\label{fig:Period}
\end{center}
\end{figure}

The equations of motion (\ref{eq:xmDot4})-(\ref{eq:mu11Dot4}) become a
five-dimensional   
dynamical system of $\vec{\xi}$=$\begin{pmatrix}
\xm & \ym & \mu_{2,0} &  \mu_{0,2} & \mu_{1,1}\end{pmatrix}$.  A
standard linear 
stability analysis yields a Jacobian:
\begin{equation}  
\begin{pmatrix}
\lambda_1 & -1 & 0 & 0 & 0 \\
k & 0 & 0 & 0 & 0 \\
0 & 0 & \lambda_2 & 0 & 0 \\
0 & 0 & 0 & 0 & 2k \\
0 & 0 & k & -1 & \lambda_3
\end{pmatrix}
\label{eq:Jacobian}
\end{equation}
where
\begin{eqnarray}
\lambda_1 &=& \sigma^2 - 1 +
3(\sigma^2-2)\mu^*_{2,0}-15(\mu^*_{2,0})^2 \\ 
\lambda_2 &=& 2(2\sigma^2 - 1-D)+6(2\sigma^2-4)\mu^*_{2,0}-90(\mu^*_{2.,0})^2\\
\lambda_3 &=& \sigma^2-1-D + 3(\sigma^2-2)\mu^*_{2,0} - 15 (\mu^*{2,0})^2.
\end{eqnarray}
The Jacobian (\ref{eq:Jacobian}) is in a block diagonal form
and the stability of
$\xm$ and $\ym$ are separated from that of the higher order moments,
which 
makes analytical investigation easier.
The 2-by-2 block at the top-left corner determines the stability of
$\xm^*=\ym^*=0$ and its
eigenvalues are given by
\begin{equation}
\lambda = \frac{1}{2}\left( \lambda_1\pm\sqrt{\lambda_1^2-4k} \right),
\label{eq:Eigenvalues}
\end{equation}
which indicates that the fixed point becomes unstable at
$\lambda_1=0$.  This 
bifurcation condition
is identical to Eq.~(\ref{eq:GaussianBifurcation1}) and therefore, the
two-variable model also undergoes reentrant transition because of the same
reason as in the single variable model.
Figure \ref{fig:HopfPhase}
compares the phase boundary obtained by the Gaussian approximation and
the simulation results.  Quantitatively, the disagreement is rather
large.  However, the qualitative features are correctly captured.

Since the eigenvalue has an imaginary part at the bifurcation point,
it is a Hopf bifurcation and a stable Limit cycle is formed above the
critical noise intensity, which is
confirmed by numerical simulation. Figure~\ref{fig:Simulation}
shows snapshots of an ensemble of particles. One observes a quite
regular limit cycle motion of the mean, even though the system is
rather small ($N=625$) and the individual units are spread widely
around the mean. The width of the cloud in $x$ direction is much
larger than the one in $y$ direction due to small $k$, which is in a
good agreement with Eq.~(\ref{eq:EquilibriumMu02}). 
Analogous to the single-variable case we will
call the oscillating phase the ordered phase and the non-oscillating
phase the disordered phase.

From the imaginary part of the eigenvalue
the period of oscillation near the bifurcation point is
approximately given by $T = 2\pi/\sqrt{k}$.  Since it
does not depend on any moment, this period is valid even without the
Gaussian approximation.  Indeed it perfectly agrees with numerical
simulation as shown in Fig.~\ref{fig:Period}.

\begin{figure}[t]
\begin{center}
\includegraphics[width=2.5in]{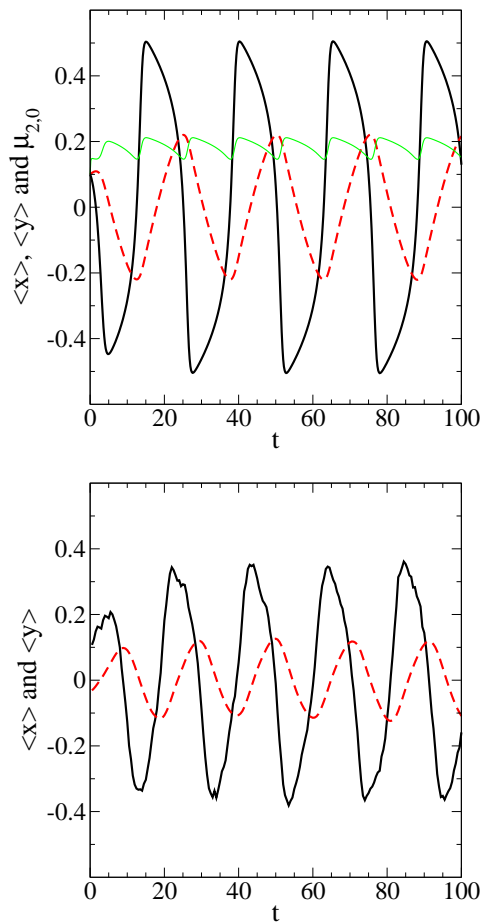}
\caption{Time evolution of $\xm$ (solid line) and $\ym$ (dashed line) obtained by numerically
  solving (\ref{eq:ymDot3}) and
  (\ref{eq:xmDot4})-(\ref{eq:mu11Dot4}) in the Gaussian
  approximation (upper panel) and 
 Langevin equations (\ref{eq:Langevin2}) and (\ref{eq:sdex})  (lower panel).
For the Gaussian approximation  $\mu_{2,0}$ (thin line) is also shown. Parameters $D=10$,
  $\sigma^2=2$ and $k=0.1$ are used.}
\label{fig:Trajectories}
\end{center}
\end{figure}

If the time evolution of the means is needed, Langevin equation or
time-dependent Fokker-Planck equation are usually solved numerically.
Numerical simulation of coupled Langevin equations is computationally rather
time consuming especially near the bifurcation point due to the finite
size effect.  One needs a large number of samplings to
obtain a 
reasonable statistics.  Furthermore, ensemble averaging is cumbersome 
since each realization oscillates in a different phase.  Numerical
integration of time-dependent Fokker-Planck equations does not have
a problem of statistical error but often
suffers from numerical instability.  A special care may be needed.
While it is difficult to solve moment dynamics (\ref{eq:xmDot4}),
(\ref{eq:ymDot3}) and (\ref{eq:mu20Dot4})-(\ref{eq:mu11Dot4})
analytically even with Gaussian approximation, it is much easier to
solve them numerically compared to the
Langevin or Fokker-Planck equation.  Since the moment dynamics is
deterministic, there is no 
statistical error. The upper panel of Fig.~\ref{fig:Trajectories}
shows the time-evolution of 
$\xm$, $\ym$, and $\mu_{2,0}$ with the Gaussian approximation.  Other
moments, $\mu_{0,2}$ and $\mu_{1,1}$ (not shown) converge to
stationary values.  The lower panel shows the results of numerical
simulation.  Only one realization with 10000 particles is
shown. Although the amplitude and period are overestimated by the
Gaussian 
approximation, all qualitative features are captured.
%

\begin{figure}[b]
\begin{center}
\includegraphics[width=2.5in]{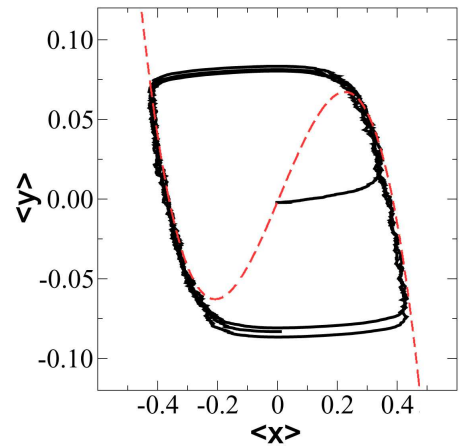}
\caption{A noise-induced relaxation oscillation.  Dashed line: a
nullcline 
determined by Eq. (\ref{eq:nullcline}).  Solid line: a limit cycle
trajectory obtained by numerical simulation. Parameter values
are $D$=10, $\sigma^2$=2, and $k$=0.1.  The simulation result follows 
the theoretical nullcline and jumps to another branch of the nullcline.}  
\label{fig:nullcline}
\end{center}
\end{figure}

Finally, we discuss a special case where $k \ll 1$ (relaxation
oscillation limit).  In this limit, Eq.~(\ref{eq:ymDot3}) indicates
$\ym$ varies very slowly.   Furthermore,
from Eq.~(\ref{eq:EquilibriumMu02})
the fluctuation of the variable $y$ is negligibly
small, suggesting that all particles experience the same value of
$y$. Since $x$ dynamics is much faster than $y$, a ``stationary''
probability distribution is formed before $y$ varies. In another word,
the variable $y$ in Eq.~(\ref{eq:xmDot3}) is just a parameter for the
dynamics of $x$.  In this case, we can investigate the Hopf
bifurcation using 
the self-consistent equation (\ref{eq:SelfCon}).
Following Ref.~\cite{vandenbroeck94}, the stationary distribution is
given by 
\begin{widetext}
\begin{equation}
P^{st}(x;\xm,\ym)=
\frac{1}{Z}\exp\left[ \frac{2}{\sigma^2}
\int_0^x dx' \left \{ -x' -\frac{\sigma^2}{2}\frac{2x'}{1+x'^2}
-\frac{D(x'-\xm)+\ym}{(1+x'^2)^2} \right\} \right]
\end{equation}
where $Z$ is a normalization constant. The self-consistent 
equation (\ref{eq:SelfCon})
yields a relation between $\xm$ and $\ym$:

\begin{equation}
\xm = \frac{1}{Z}\int_{-\infty}^{+\infty} dx \, \frac{x}{1+x^2}
\exp \left[ -\frac{x^2}{\sigma^2} +
\frac{D}{\sigma^2}\frac{1}{1+x^2}
+ \frac{D\xm-\ym}{\sigma^2}
\left\{ \frac{x}{1+x^2}+\arctan x\right\} \right]
\label{eq:nullcline}
\end{equation}
\end{widetext}
which corresponds to a nullcline of
the relaxation oscillation. Figure~\ref{fig:nullcline}
plots the nullcline and an actual
trajectory from numerical simulation.  The trajectory follows one
branch of the nullcline and jumps to the other branch analogous to the
deterministic relaxation oscillation.  This nullcline is exact.
However, its derivation requires an analytical expression of the
probability distribution and may not be applicable to other cases.
An alternative method such as the Gaussian
approximation is useful for general cases where the analytical
distribution is not available.
%

\section{Conclusions}

We have described a general theoretical method useful for the 
investigation of 
nonlinear stochastic systems exhibiting noise-induced phase
transitions. It is based on the dynamical equations for the
central moments. In general the exact solution involves the solution
of the infinite dimensional set of ordinary differential equations for
these moments. However, even without solving this system useful
qualitative 
information can be gathered from the equations. Furthermore,
quantitatively reasonable solutions can be obtained by neglecting the
cumulants of 
the distributions above a certain order and then solving the remaining
finite set of equations.

Using this method we investigated two systems. One exhibits a noise
induced pitchfork bifurcation, the other one a Hopf bifurcation. In
the latter case, the macroscopic quantities oscillate in time when the
system is in an ordered phase.  This oscillation is purely induced by
noise via spontaneous symmetry breaking. The macroscopic oscillation
suggests a strong synchronization of microscopic degrees of freedom
despite the presence of noise. Actually, it is the noise that
generates the macroscopic order. On the other hand a strong noise
destroys the order again. The reentrance into the disordered phase is
due to the fourth moment, that grows faster with noise intensity than
the second moment.

In the Gaussian approximation we have reproduced the basic features of
the noise-induced phase transition, namely the existence of a critical
coupling strength and the disorder-order-disorder transition.

\acknowledgments
This work was supported in part by the National Science Foundation
under Grant Nos. PHY-9970699 and DMS-0079478, the ESF program STOCHDYN
and the ''Deutsche Forschungsgemeinschaft'' under the
Sonderforschungsbereich 555 ''Komplexe Nichtlineare Prozesse''. 

\bibliographystyle{apsrev}

\end{document}